\documentclass[preprint,showpacs,amsmath,amssymb]{revtex4}

\usepackage{graphicx}
\usepackage{dcolumn}
\usepackage{bm}

\begin{document}

\title{Strangeness Production Process $pp \rightarrow nK^ +  \Sigma ^ +$ \\within Resonance Model}

\author{Xu Cao$^{1,3}$, Xi-Guo Lee$^{1,2}${\footnote{Email: xgl@impcas.ac.cn}}, and Qing-Wu Wang$^{1,3}$}

\affiliation{1. Institute of Modern Physics, Chinese Academy of
Sciences, P.O. Box 31, Lanzhou 730000, P.R.China\\2. Center of
Theoretical Nuclear Physics, National Laboratory of Heavy Ion
Collisions, Lanzhou 730000, P.R.China\\3. Graduate School, Chinese
Academy of Sciences, Beijing 100049, P.R.China}


\begin{abstract}

We explore production mechanism and final state interaction in the
$pp \rightarrow nK^ + \Sigma ^ +$ channel based on the inconsistent
experimental data published respectively by COSY-11 and COSY-ANKE.
The scattering parameter $a> 0$ for $n\Sigma ^ +$ interaction is
favored by large near-threshold cross section within a
nonrelativistic parametrization investigation, and a strong $n\Sigma
^ +$ interaction comparable to $pp$ interaction is also indicated.
Based on this analysis we calculate the contribution from resonance
$\Delta ^* (1920)$ through $\pi ^ +$ exchange within resonance
model, and the numerical result suggests a rather small
near-threshold total cross section, which is consistent with the
COSY-ANKE data. With an additional sub-threshold resonance $\Delta
^*(1620)$, the model gives a much better description to the rather
large near-threshold total cross section published by COSY-11.

\end{abstract}

 \maketitle{}

\section{INTRODUCTION}

Strangeness production process in nucleon-nucleon collisions has
attracted considerable theoretical interest since high precision
data have been published in the past few
years\cite{1,2,3,4,5,6,7,8}. This field is fascinating for several
reasons. First of all, strangeness production in nucleon-nucleon
collisions is an elementary process, and their cross sections are an
fundamental input into transport model calculations of the
strangeness production in heavy ion collisions. Furthermore,
strangeness production process may provide information on the
strange components of the nucleon and deepen our knowledge on the
internal structure of nucleon\cite{22}. Finally, for the short life
time of hyperons, it is difficult to accumulate a large set of
scattering events and obtain accurate scattering parameters. Final
state interaction (FSI) in strangeness production can supply us with
assistant information on the KN and YN interaction.

A series of theoretical models\cite{9,10,11,12,13} have given
excellent description to the observed energy dependence of the total
cross section for the strangeness production process $pp \rightarrow
pK^ +  \Lambda$ and $pp \rightarrow pK^ +  \Sigma ^ 0$. Some of
these models\cite{11,12,14} also give a reasonable explanation to
the energy dependence of the ratio $R={\sigma (pp \rightarrow pK^ +
\Lambda )/\sigma (pp \rightarrow pK^ +  \Sigma ^0 )}$. Compared to
the detailed experimental and theoretical investigation in these two
channels, there is only limited data on the total cross section for
$pp \rightarrow nK^ +  \Sigma ^ +$. A recent COSY-ANKE data\cite{26}
indicates a rather small near-threshold total cross section, which
is much smaller than previous COSY-11 data\cite{25}. For the large
error bar of these experimental data, any definite conclusion cannot
be surely drawn at present, and it must be checked in future
experiments.

On the other hand, as a good isospin-3/2 filter without complication
caused by $N^{*}$ contribution, $pp \rightarrow nK^ +  \Sigma ^ +$
is an excellent channel for the investigation of $\Delta^{*}$
resonance\cite{17}. It is necessary to deepen our theoretical
understanding to the controversial experimental data. In this paper,
we use a resonance model and its nonrelativistic parametrization to
explore the energy dependence of the total cross section of the $pp
\rightarrow nK^ + \Sigma ^ +$. In section II we introduce the
Feynman diagrams and the effective Lagrangian approach within
resonance model. In section III we present a exploration to the near
threshold region within a nonrelativistic parametrization and give a
set of FSI parameters. In section IV, we give the numerical results
and explore their implications, especially in the near threshold
region. Section V provides a summary.

\section{Resonance Model and Effective Lagrangian Approach}

The treatment of the elementary processes $NN \rightarrow NYK$ is
tree level at the hadronic level in the resonance model\cite{9}, as
illustrated in Figure 1. All interference terms between different
amplitudes are neglected because the relative phases of these
amplitudes can not be fixed by the scarce experimental data. Mesons
exchanged are restricted to those observed in the decay channels of
the adopted resonances, and kaon exchange is not included in the
calculation. Thus, all values of the coupling constants are fixed by
the experimental decay ratios, and the only adjustable parameters
are cut-off parameters in the form factors. In this model, only
$\Delta^{*}(1920)$ through $\pi ^ +$ exchange contributes to the $pp
\rightarrow nK^ +  \Sigma ^ +$ channel above the production
threshold. The relevant meson-nucleon-nucleon(MNN) and
resonance-nucleon(hyperon)-meson effective Lagrangians for
evaluating the Feynman diagrams in Fig. 1 are:

\begin{equation}
L_{\pi NN}  =  - ig_{\pi NN} \bar N\gamma _5 \vec \tau  \cdot \vec
\pi N ,
\end{equation}

\begin{equation}
L_{\Delta (1920)N\pi }  = \frac{{g_{\Delta (1920)N\pi } }}{{m_\pi
}}\bar N\Delta _\mu  \vec \tau  \cdot \partial ^\mu \vec \pi  + h.c.
,
\end{equation}

\begin{equation}
L_{\Delta (1920)\Sigma K}  = \frac{{g_{\Delta (1920)\Sigma K}
}}{{m_K }}\Delta _\mu  \vec I  \cdot \vec \Sigma
\partial ^\mu  K + h.c. ,
\end{equation}
where $N$,$\Delta _\mu$,$K$ and $\vec \pi$ are the Rarita-Schwinger
spin wave function of the nucleon, the resonance, the kaon and the
pion respectively, $\vec \tau$ the Pauli matrices, $\vec I$ the
spin-3/2 matrices and $g_{\pi NN}^2 /4\pi = 14.4$. The effective
Lagrangians for the resonance $N^{*}$ please refer to
\cite{9,15,16,17,18}.

For the propagators of $\pi$ meson and the spin-3/2
$\Delta^{*}(1920)$ resonance, the usual Rarita-Schwinger propagators
are used:

\begin{equation}
G_\pi  (k_\pi  ) = \frac{i}{{k_\pi ^2  - m_\pi ^2 }} ,
\end{equation}

\begin{equation}
G_R^{\mu \nu } (p_R ) = \frac{{\gamma \cdot p + m_R }}{{p_R^2  -
m_R^2 + im_R \Gamma _R }}[g_{\mu \nu }  - \frac{1}{3}\gamma _\mu
\gamma _\nu   - \frac{1}{{3m_R }}(\gamma _\mu  p_\nu   - \gamma _\nu
p_\mu ) - \frac{2}{{3m_R^2 }}p_\mu  p_\nu  ] ,
\end{equation}
with $k_\pi$ and $p_R$ being the four momentum of $\pi$ meson and
$\Delta^{*}(1920)$ resonance respectively, $m_\pi$ and $m_R$ the
corresponding masses. The relations between the branching ratios and
the coupling constants then can be calculated using the relevant
Lagrangians. Take $\Delta^{*} (1920) \to N\pi$ as an example:

\begin{equation}
\Gamma _{\Delta (1920)N\pi }  = \frac{{g_{\Delta (1920)N\pi }^2
}}{{m_\pi ^2 }}\frac{{(m_N  + E_N )(P_N^{cm} )^3 }}{{12\pi M_{\Delta
(1920)} }} ,
\end{equation}

\begin{equation}
P_N^{cm}  = \frac{{\lambda ^{1/2} (m_{\Delta (1920)}^2 ,m_N^2 ,m_\pi
^2 )}}{{2M_{\Delta (1920)} }} ,
\begin{array}{*{20}c}
   {} & {}  \\
\end{array}
E_N  = \sqrt {(P_N^{cm} )^2  + m_N^2} ,
\end{equation}
here $\lambda (x,y,z) = x^2  + y^2  + z^2  - 2xy - 2yz - 2zx$. Then
the coupling constants can be determined through the empirical
branching ratios. Values related to $\Delta^{*}(1920)$ are
summarized in Table I\cite{9,17}.

The resulting $\pi NN$ and $\Delta (1920)N\pi$ vertexes are
multiplied by form factors which dampen out high values of the
exchanged momentum:

\begin{equation}
F_M (\vec q) = \frac{{\Lambda _M^2  - m_M^2 }}{{\Lambda _M^2  -
q_M^2 }} ,
\end{equation}
with $\Lambda _M$, $q_M$ and $m_M$ being the cut-off parameter,
four-momentum and mass of the exchanged meson. $\Lambda _\pi =
1.3GeV$ as the widely used value\cite{9,15,17}. The form factor of
the $\Delta (1920)\Sigma K$ vertex is 1.

\begin{table}
\caption{Relevant $ \Delta ^* (1920)$ parameters} \label{table1}
\begin{center}
\begin{tabular}{c  c c c c c c c}
\hline\hline
 &  & Width & Channel & Branching ratio & Adopted value & $g^2 /4\pi $ &
\\
\hline & $\Delta ^* (1920)$ &200$MeV$ & $N\pi$ & 0.05-0.20 & 0.125 &
0.0034 &
\\
 &   &   & $\Sigma K$ & 0.01-0.03&  0.021 & 0.0292  &\\
\hline \label{table1}
\end{tabular}
\end{center}

\end{table}

As to $pp \rightarrow nK^ +  \Sigma ^ +$ channel, if the only
contribution comes from resonance $\Delta^{*}(1920)$ through $\pi ^
+$ exchange, the corresponding amplitude can be obtained
straightforwardly by applying the Feynman rules to Fig. 1:

\begin{equation}
{\rm M} = \frac{{g_{\Delta (1920)\Sigma K} }}{{m_K }}\bar
u_{(p_\Sigma  )}^{(S_\Sigma  )} p_K^\mu  G_{\Delta (1920)}^{\mu \nu
} p_{\Delta (1920)} \frac{{g_{\Delta (1920)N\pi } }}{{m_\pi
}}u_{(p_1 )}^{(S_1 )} p_\pi ^\nu  G_\pi  (p_\pi  )\bar u_{(p_n
)}^{(S_n )} \sqrt 2 g_{\pi NN} \gamma _5 u_{(p_2 )}^{(S_2 )}
\nonumber
\end{equation}
\begin{equation}
 +(exchange
\begin{array}{*{20}c}
   {} & {}  \\
\end{array}
 term
\begin{array}{*{20}c}
{} & {}  \\
\end{array}
with
\begin{array}{*{20}c}
{} & {}  \\
\end{array}
p_1 \leftrightarrow p_2) ,
\end{equation}

Final state interaction influences the near-threshold behavior
significantly. It has been experimentally and theoretically verified
that $p\Lambda$ FSI is essential to the $pp \rightarrow pK^ +
\Lambda$ process. Similarly, here only $n\Sigma ^ +$ interaction is
considered with Watson-Migdal factorization\cite{19}, and $ nK^ + $
interaction will not be taken into account for its weakness\cite{8}:

\begin{equation}
{\rm A} = {\rm M}T_{n\Sigma } ,
\end{equation}
where $T_{n\Sigma }$ is the enhancement factor describing the
$n\Sigma ^ +$ final state interaction and goes to unity in the limit
of no FSI, as expected by the physical picture. Similar to the
$p\Lambda$ FSI in the $pp \rightarrow pK^ +  \Lambda$ channel, we
assume that $T_{n\Sigma }$ can be taken to be Jost
function\cite{20}:

\begin{equation}
T_{n\Sigma }  = \frac{{q + i\beta }}{{q - i\alpha }} ,
\begin{array}{*{20}c}
{} & {}  \\
\end{array}
\end{equation}

\begin{equation}
q =\frac{{\lambda ^{1/2} (s,m_\Sigma ^2 ,m_n^2 )}}{{2\sqrt s }} ,
\begin{array}{*{20}c}
{} & {}  \\
\end{array}
s=(\varepsilon  + m_n  + m_{\rm K}  + m_\Sigma  )^2 ,
\end{equation}
with $\varepsilon$ being the excess energy. The related scattering
length and effective range are:

\begin{equation}
a = \frac{{\alpha  + \beta }}{{\alpha \beta }} ,
\begin{array}{*{20}c}
   {} & {}  \\
\end{array}
r =\frac{2}{{\alpha  + \beta }} \label{eq:fsia} ,
\end{equation}

Then the total cross section can be calculated by:

\begin{equation}
\sigma _{tot}  = \frac{{m_p^2 }}{{4F}}\int {\left| {\rm M} \right|}
^2 \delta ^4 (p_1  + p_2  - p_n  - p_K  - p_\Sigma  )\frac{{m_n d^3
p_n }}{{E_n }}\frac{{d^3 p_K }}{{2E_K }}\frac{{m_\Sigma  d^3
p_\Sigma  }}{{E_\Sigma  }} \label{eq:phase} ,
\end{equation}
with the flux factor $F = (2\pi )^5 \sqrt {(p_1  \cdot p_2 )^2  -
m_p^4 }$. $\left| {\rm M} \right|^2$ is the square of the invariant
scattering amplitude, averaged over the initial spins and summed
over the final spins. The integration over the phase space can be
performed by Monte Carlo program.

For the lack of the $n\Sigma ^ +$ scattering data, a set of data are
employed referring to the $p\Lambda$ interaction\cite{21} in the
Ref.\cite{17}'s calculation:

\begin{equation}
\alpha  =  - 75.0MeV ,
\begin{array}{*{20}c}
{} & {}  \\
\end{array}
\beta  = 200.0MeV ,
\end{equation}

The corresponding scattering length and effective range are:

\begin{equation}
a  =  - 1.6fm ,
\begin{array}{*{20}c}
{} & {}  \\
\end{array}
r  = 3.2fm , \label{eq:fsib}
\end{equation}

Next we will concentrate on the relation between FSI and the
near-threshold total cross section, and try to find some clues on
the $n\Sigma ^ +$ interaction. As a matter fact, we fix scattering
length and effective range with a nonrelativistic parametrization,
assuming no possible quasi-bound or molecular state.

\section{Nonrelativistic Parametrization and FSI}

It is generally agreed that the energy variation of the total cross
section for the $pp \rightarrow pK^ +  \Lambda$ and $pp \rightarrow
pK^ +  \Sigma ^ 0$ is fixed mainly by the $\varepsilon ^2$ factor
coming from phase space, modified by the FSI in the near-threshold
region:

\begin{equation}
\sigma _{tot}  \propto \varepsilon ^2  \times FSI ,
\end{equation}

If $\left| {\rm M} \right|^2$ depend smoothly on the excess energy,
Eq.\eqref{eq:phase} can be parameterized as\cite{21,23}:

\begin{equation}
\sigma _{tot}  = \frac{{\sqrt {m_N m_{\rm K} m_\Sigma  } }}{{2^7 \pi
^2 (m_N  + m_{\rm K}  + m_\Sigma  )^{3/2} }}\frac{{\varepsilon ^2
}}{{\lambda ^{1/2} (s,m_N^2 ,m_N^2 )}}\left| M \right|^2 \kappa
,\label{eq:pp}
\end{equation}

\begin{equation}
\kappa  = 1 + \frac{{4\beta ^2  - 4\alpha ^2 }}{{( - \alpha  + \sqrt
{\alpha ^2  + 2\mu \varepsilon } )^2 }} ,
\end{equation}
where $\kappa$, the FSI factor, was firstly put forward by G.
F$\ddot{a}$ldt and C. Wilkin\cite{28}. At sufficiently large
$\varepsilon$, $\kappa \rightarrow 1$ and FSI can be isolated. It is
noted that the nonrelativistic phase space factor is used in the
above equation, and this is a good approximation up to excess
energies of $\epsilon \simeq 1GeV$. In order to give some clues to
the $n\Sigma ^ +$ interaction, the physical mechanism of strangeness
production will not be concerned, and $\left| {\rm M} \right|^2$ is
parameterized as $\left| {\rm M} \right|^2  \propto e^{ -
const.\varepsilon }$ . Similar parametrization\cite{21} has give an
excellent description to the energy dependence of the total cross
section for the $pp \rightarrow pK^ +  \Lambda$, $pp \rightarrow pK^
+  \Sigma ^ 0$ and the ratio $R={\sigma (pp \rightarrow pK^ +
\Lambda )/\sigma (pp \rightarrow pK^ +  \Sigma ^0 )}$.

If parameters in FSI factor are employed to be the same with
Eq.\eqref{eq:fsib}, and $\left| {\rm M} \right|^2 = 18.9e^{ -
1.3\varepsilon }  \cdot 10^6 \mu b$  ($\varepsilon$ in GeV), the
total cross section for $pp \rightarrow nK^ +  \Sigma ^ +$ as a
function of excess energy can be exhibited by the solid curve in
Fig. 2. It gives a reasonable description to the total cross section
in high energies and underestimates the COSY-11 data, which is
consistent with the numerical calculation below(see Fig.3). This
verifies the applicability of the parametrization Eq.\eqref{eq:pp}.

If $n\Sigma ^ +$  FSI is negligible ($\kappa  = 1$) and $\left| {\rm
M} \right|^2  = 21.2e^{ - 1.3\varepsilon }  \cdot 10^6 \mu b$ , the
result is dashed curve in Fig. 2. It still overestimates COSY-ANKE
data by a factor of about 3. As a matter of fact, the pure phase
space behavior is favored by so small near-threshold total cross
section, as indicated by the dotted curve with $\kappa = 1$ and
$\left| {\rm M} \right|^2  = 4.7 \cdot 10^6 \mu b$ . This means that
$n\Sigma ^ +$  FSI is negligible, similar to the $p\Sigma ^0$ FSI in
$pp \rightarrow pK^ +  \Sigma ^ 0$,  but the scattering amplitude is
more weakly dependent on the excess energy. If the near-threshold
total cross section is really as small as COSY-ANKE data\cite{26},
the nearly constant scattering amplitude is the extraordinary
character for $pp \rightarrow nK^ +  \Sigma ^ +$ channel.

However, in order to reproduce COSY-11 data, totally different
parameters should be adopted:

\begin{equation}
\left| {\rm M} \right|^2  = 20e^{ - 1.7\varepsilon }  \cdot 10^6 \mu
b ,
\end{equation}

\begin{equation}
\alpha  = 30MeV ,
\begin{array}{*{20}c}
{} & {}  \\
\end{array}
\beta  = 360MeV  ,\label{eq:new}
\end{equation}

The result is the dash-dotted curve in Fig. 2. The corresponding
scattering length $a = 7.1fm > 0$  (with $r = 1.0fm$ )  is contrary
to the values used in the Ref.\cite{17}'s calculation
(Eq.\eqref{eq:fsib}) and implies an even stronger $n\Sigma ^ +$
interaction. It is worth noting that the scattering parameters of
$pp$ interaction are $a = -7.8fm $ and $r = 2.8fm$. In the next
section we will give a comparison between these two sets of $n\Sigma
^ +$ FSI parameters, and show that a better reproduce to the large
near-threshold total cross section is achieved with
Eq.\eqref{eq:new}.

\section{Numerical Results and Implication}

The contribution from resonance $\Delta^{*}(1920)$ through $\pi ^ +$
exchange with FSI parameters in Eq.\eqref{eq:fsib} is shown as the
dotted curve in Fig. 3. Note that the resonance $\Delta^{*}(1920)$
can be treated as an effective resonance which represents all
contributions of six resonances
($\Delta^{*}(1900)$,$\Delta^{*}(1905)$,$\Delta^{*}(1910)$,$\Delta^{*}(1920)$,$\Delta^{*}(1930)$,
and $\Delta^{*}(1940)$) and the coupling constants relevant to
$\Delta^{*}(1920)$ are scaled multiplying by a factor\cite{27}. The
numerical result (the dashed curve in Fig. 3) reproduces the data in
high energies reasonably well but underestimates COSY-11 data by a
factor of orders, as pointed out in above section. It has been
suggested\cite{17} that the inclusion of a ignored sub-threshold
resonance $\Delta^{*}(1620) 1/2^{-}$ (both $\rho^{+}$ and $\pi ^ +$
exchange are included), together with a strong $n\Sigma ^ +$ FSI in
Eq.\eqref{eq:fsib}, would achieve much better agreement, as
illustrated by the solid curve in Fig. 3. Unfortunately, it still
underpredicts two near-threshold data points. It is argued\cite{17}
that improvement can be achieved by inclusion of the interference
term between $\rho^{+}$ and $\pi ^ +$ exchange. However, this kind
of interference term is very small as demonstrated in another
channel $pp \rightarrow pK^ +  \Lambda$\cite{16}, and it definitely
can not provide any interpretation to this relatively large
discrepancy.

The numerical result showed by the dashed curve in Fig. 3 agrees
COSY-ANKE data, which was obtained at a little higher energy and
shows a rather small total cross section in near-threshold region,
as indicated by a closed star in Fig. 3.

The numerical results with the FSI parameters in Eq.\eqref{eq:new}
are given in Fig.4. The contribution from resonance
$\Delta^{*}(1920)$ through $\pi ^ +$ exchange is not affected much
by the alteration of the FSI parameters, and still gives a small
near-threshold total cross section. This indicates that the
contribution from resonance $\Delta^{*}(1920)$ is not sensitive to
the magnitude of the $n\Sigma ^ +$ interaction, and we would be not
able to distill any valuable information about the $n\Sigma ^ +$
interaction. However, the contribution from resonance
$\Delta^{*}(1620)$ gives an even larger near-threshold cross
section, and the resulting curve reproduces all previous published
data reasonably, as illustrated by the solid curve in Fig. 4. This
strongly suggests a rather strong $n\Sigma ^ +$ interaction if the
near-threshold cross section is as large as COSY-11 data\cite{25}.
We would like to stress that this is incompatible to the known YN
scattering data. Besides, the $p \Sigma ^ 0$ interaction is found to
be weak in $pp \rightarrow pK^ + \Sigma ^ 0$ channel, and the naive
speculation is that the $n\Sigma ^ +$ interaction should be weak,
too. One may further expect that the energy dependence of the $pp
\rightarrow nK^ + \Sigma ^ +$ is approximately the same with that of
$pp \rightarrow pK^ +  \Sigma ^ 0$\cite{29}. Anyway, due to the
scarcity and uncertainty of those experimental data, the conclusion
that these data imply a highly anomaly near-threshold behavior can
not be definitely drawn yet. Based on these numerical result, we may
also conclude that contribution from $\Delta^{*}(1620)$ is
indispensable for a good reproduce to the large near-threshold total
cross section, which has been discussed deeply in the Ref.\cite{17}.

\section{Summary}

We presented the confusion caused by the existing data for $pp
\rightarrow nK^ +  \Sigma ^ +$ channel. The numerical result and
parametrization analysis reveal a large discrepancy between two sets
of near-threshold data, which means totally different $n\Sigma ^ +$
interaction. The theoretical results presented in other
works\cite{21} also indicated that the total cross section data
alone cannot distinguish different production mechanisms. In order
to clarify the confusion that if there was a highly abnormal
near-threshold behavior in $pp \rightarrow nK^ +  \Sigma ^ +$
channel, it is necessary to accumulate a data set with high accuracy
and large statistics. Further measurements are being planned in
COSY-ANKE\cite{29}. Besides, a significant improvement is to be
expected through the installation of Cooling Storage Ring (CSR) in
Lanzhou, China, which is designed for the study of heavy-ion
collisions. It can provide $1 \sim 2.8GeV$ proton beam and perform
accurate measurement to differential observables and invariant mass
spectrum. In Ref.\cite{21} the authors proposed a scheme to study
the role of the FSI by analyzing Dalitz plot distribution, and
experiment performed by COSY-TOF Collaboration\cite{8} verified
their prediction. Similar analysis may also clarify the confusion in
$pp \rightarrow nK^ + \Sigma ^ +$ channel. So CSR 's data may offer
an opportunity to explore the reaction mechanism for the strange
production process in nucleon-nucleon collisions. Only then it is
possible to achieve an acceptable confidence level for the
near-threshold behavior and the dynamical mechanism in the $pp
\rightarrow nK^ +  \Sigma ^ +$ channel.

\begin{acknowledgments}

We would like to thank J. J. Xie and B. S. Zou for fruitful
discussions and program code. We also thank Colin Wilkin for a
careful reading of the draft and valuable comments. This work was
supported by the CAS Knowledge Innovation Project
(No.KJCX3-SYW-N2,No.KJCX2-SW-N16) and Science Foundation of China
(10435080, 10575123,10710172).

\end{acknowledgments}

\begin{figure}[ht]
  \begin{center}
    \rotatebox{0}{\includegraphics*[width=0.7\textwidth]{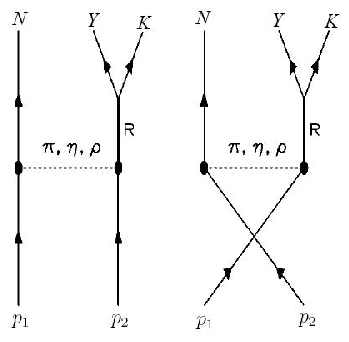}}
    \caption{Feynman diagrams in the resonance model. $N$, $Y$ and $K$
stand for, respectively, the nucleon, the hyperon and the kaon. $R$
is the resonance.}
  \end{center}
\end{figure}

\begin{figure}[ht]
  \begin{center}
    \rotatebox{0}{\includegraphics*[width=0.7\textwidth]{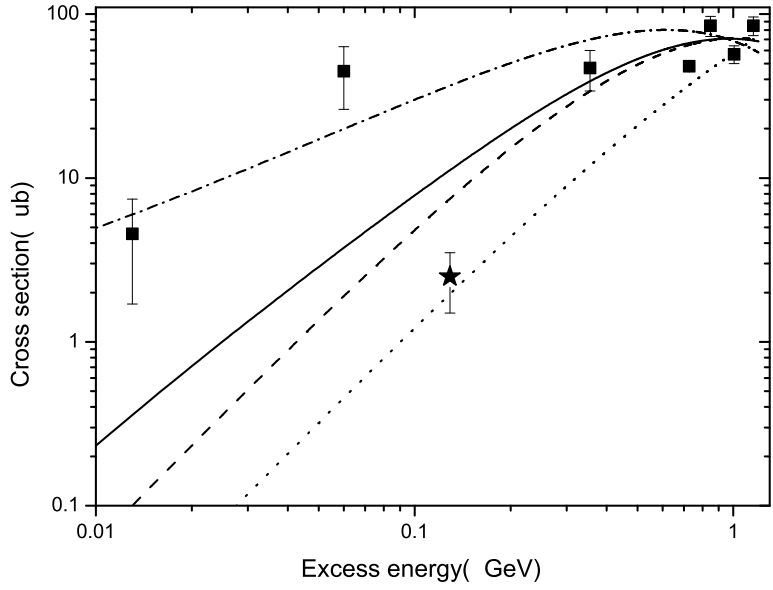}}
  \caption{Total cross section for $pp \rightarrow nK^ +  \Sigma ^ +$
as a function of excess energy. The relevant parameters used for the
curves refer to the text. Data are taken from Ref.\protect\cite{24},
COSY-11\protect\cite{25}(closed square) and
COSY-ANKE\protect\cite{26} (closed star).}
  \end{center}
\end{figure}

\begin{figure}[ht]
  \begin{center}
    \rotatebox{0}{\includegraphics*[width=0.7\textwidth]{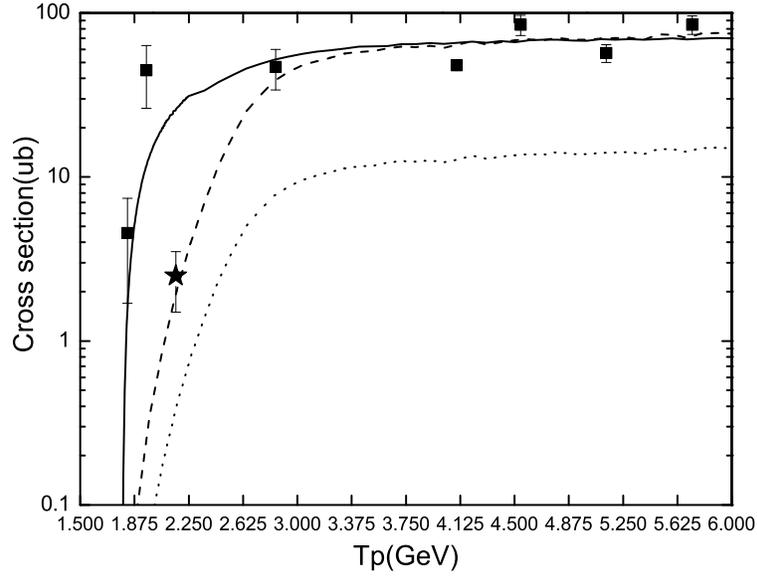}}
  \caption{Variation of the total cross section with the kinetic
energy of the proton beam (Tp). Dotted curve: contribution from
resonance $\Delta ^* (1920)$ through $\pi ^ +$ exchange. Dashed
curve: contribution from resonance $\Delta ^* (1920)$ scaled
multiplying by a factor 5. Solid curve: numerical result in the
Ref.\protect\cite{17}. Data are the same as in Figure 2.}
  \end{center}
\end{figure}

\begin{figure}[ht]
  \begin{center}
    \rotatebox{0}{\includegraphics*[width=0.7\textwidth]{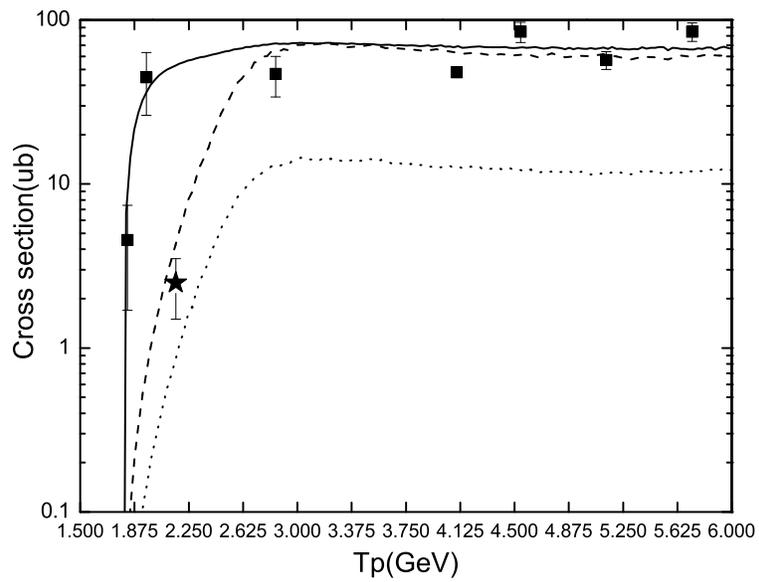}}
  \caption{Same as in Fig. 3, but with the FSI parameters in Eq.\protect\eqref{eq:new}. Data are the same as in Figure 2.}
  \end{center}
\end{figure}

\end{document}